\documentclass[letterpaper,draftcls,onecolumn,journal,12pt]{IEEEtran} 
\usepackage[dvips]{graphicx,color}
\usepackage{amsmath,amssymb,cite,bm,balance,cases,multirow}
\usepackage{citesort} 
\usepackage{alltt}
\usepackage{mediabb}

\usepackage{algorithm,algorithmic}
\renewcommand{\algorithmicrequire}{\textbf{Input:}}
\renewcommand{\algorithmicensure}{\textbf{Output:}}

\usepackage{multirow}
\usepackage[normalem]{ulem}

\begin{document}

\title{A New Polar Code Design Based on \\
Reciprocal Channel Approximation}

\author{Hideki Ochiai, 
Kosuke Ikeya,
and Patrick Mitran 
\thanks{H. Ochiai and K. Ikeya are with the Department of Electrical and Computer Engineering, 
Yokohama National University, Yokohama, Japan. (email: hideki@ynu.ac.jp, ikeya-kosuke-fc@ynu.jp)}
\thanks{P. Mitran is with the Department of Electrical and Computer Engineering, 
University of Waterloo, ON, Canada. email:pmitran@uwaterloo.ca}
}
\date{}

\maketitle
\quad \vspace{-3.2cm}\\

\begin{abstract}
This paper revisits polar code design for
a binary-input additive white Gaussian noise~(BI-AWGN) channel
when successive cancellation~(SC) decoding is applied at the receiver.
We focus on the {\em reciprocal channel approximation~(RCA)}, 
which is often adopted
in the design of 
low-density parity-check~(LDPC) codes. 
In order to apply RCA to polar code design for
various codeword lengths,
we derive rigorous closed-form approximations
that are valid over a wide range of SNR
over an AWGN channel, for both the mutual information of BPSK signaling 
and the corresponding {\em reciprocal channel mapping}.
As a result, the computational complexity 
required for evaluating channel polarization 
is thus equivalent to that based on 
the popular Gaussian approximation~(GA) approach.
Simulation results show that the proposed polar code design based on RCA
outperforms those based on GA as well as 
the so-called improved GA~(IGA) approach,
especially as the codeword length is increased.
Furthermore, the RCA-based design yields a better block error rate~(BLER) estimate
compared to GA-based approaches.
\end{abstract}
 

\IEEEpeerreviewmaketitle

\vspace{-0.5cm}

\section{Introduction}

One of the most striking properties of
polar codes~\cite{arikan09} is their capacity approaching behavior
that is achievable with low-complexity successive cancellation~(SC) decoding. 
Specifically, for a codeword length of $N$ bits, the decoding complexity
is only $O(N\log N)$, which is significantly lower than other known 
capacity approaching codes that are available in practice.
As a result of channel polarization, 
the design of polar codes is equivalent to the identification of good channels
and bad channels,
where 
the former channels are used for information transmission
and the latter channels are left unused (i.e., {\em frozen}).
The main focus of this paper is on the design of polar codes with various codeword
lengths over a binary-input additive white Gaussian noise~(BI-AWGN) channel.

There have been various techniques proposed for polar code design over a BI-AWGN channel.
The most accurate analytical approach is the use of density evolution~\cite{richardson2001,richardson_modern_2008},
originally developed for the design of low-density parity-check~(LDPC) codes,
and its applicability to polar codes has been identified in~\cite{mori2009}.
Density evolution tracks the probability distribution of the channel or 
its log-likelihood ratio~(LLR), and in order to improve its accuracy, it should be computed with
sufficient quantization and dynamic range. Therefore, density evolution is highly 
demanding in terms of space and computational complexity,
especially when the codeword length increases.
A more tractable approach with limited space complexity
was proposed in~\cite{tal2013}.
Nevertheless, the approach involves quantization,
and thus the overall complexity depends not only on 
the codeword length but also on the required precision.
On the other hand, a significantly simpler approach is 
Gaussian approximation~(GA)~\cite{chung2001}.
Also initially developed for the design of LDPC
codes, its applicability to polar code design has been well investigated~\cite{trifonov2012}.
It has been pointed out in~\cite{ochiai2021} 
that GA with the original approximation
function developed in~\cite{chung2001} may not necessarily work accurately 
when the equivalent SNR values of the channels after polarization become low.
Consequently, a modified version, which will be referred to as 
improved GA~(IGA) in this work, 
has been proposed~\cite{ochiai2021}. 
Furthermore, there have been various design approaches proposed in the recent
literature, targeting specific polar decoding algorithms.
For example, in the case of successive cancellation list~(SCL) decoding~\cite{tal2015} or belief propagation~(BP) decoding~\cite{Arikan2008bp}, learning-based approaches 
such as genetic algorithms and  reinforcement learning 
have been proposed in~\cite{Elkelesh2019} and \cite{liao2022}, respectively.
The major limitations of these approaches
are their lack of flexibility and scalability since computationally demanding
training should be performed 
for each given combination of the code parameters such as
block length and code rate. 

In this work, we consider the application of 
 the so-called {\em reciprocal channel 
approximation~(RCA)}, introduced by Chung in~\cite{chung_diss_2000},
that is motivated
by the duality property of mutual information that holds between
a repetition code and parity-check code over a binary erasure channel~(BEC)~\cite{ashikhmin2004}.
This approach has been originally adopted in 
the design of LDPC and other related 
codes~\cite{chung_diss_2000,ten_brink2003,ten_brink2004,sharon2006,divsalar2009},
and it was numerically demonstrated in~\cite[Chapter~7]{chung_diss_2000} 
that it offers a better approximation
for density evolution 
in the case of BI-AWGN channels
than the conventional GA.
By numerical study and simulation, it will be shown that polar codes designed
based on RCA exhibit better performance than those based on (improved) GA, and
the gain becomes noticeable as the block length increases.
The superior performance improvement will be demonstrated for
polar codes with very long lengths (such as $N=2^{18}$ bits), 
as is considered
in practical optical communications~\cite{Smith2012}. (See, also \cite{mahdavifar2011},
where a polar code of length $2^{20}$ bits is considered.)

The design of polar codes based on RCA has been initially considered 
in~\cite{vakilinia2015} as well as~\cite{dosio2016}.
RCA-based designs make use of 
the mutual information 
of BPSK signaling over an AWGN channel as well
as its inverse function, which should be calculated numerically. 
There are several closed-form approximations available in the literature,
and in~\cite{dosio2016} the two approximations 
developed in~\cite{ten_brink2004} and \cite{brannstrom2005} are compared,
indicating the sensitivity of polar code performance to 
the choice of approximation.
For this reason, in this work we introduce 
rigorous closed-form expressions for the mutual information $C(\gamma)$ of BPSK 
as a function of SNR~$\gamma$
as well as its inverse function~$C^{-1}(\cdot)$ that can be used for design of polar codes
with long codeword lengths.

The main contributions of this work are as follows:
\begin{itemize}
 \item We derive a closed-form piece-wise continuous
mutual information expression
for BPSK signaling over AWGN channels with
 guaranteed convergence in the case of high and low SNR
based on the asymptotic analysis of the 
mutual information function. 
This expression is exploited to design  polar codes 
based on RCA
supporting a wide range of SNR after polarization.

\item We develop an explicit algorithm 
that identifies SNR after channel polarization
using only closed-form equations, which can thus be calculated with low complexity.

\item By simulation, we demonstrate that
the RCA-based design can achieve better performance as well as 
offer a better block error rate~(BLER) estimate
compared to GA-based approaches, especially
as the block length increases and the code rate is moderate (e.g.,
around $1/2$).
\end{itemize}

The remainder of the paper is organized as follows. 
Section~\ref{sec:Pol} reviews the principle of RCA from the viewpoint of polar code design
and develops a general algorithm. The key function
$\Lambda(\xi)$ that 
corresponds 
to the {\em reciprocal channel mapping} defined in~\cite{chung_diss_2000}
and that transforms SNR in the log domain based on RCA,
is introduced.
Section~\ref{sec:cap_expression} develops closed-form expressions of the mutual information
for BPSK signaling as well as its inverse, and its accuracy is compared with 
two other formulas available in the literature.
Based on the developed mathematical tools, closed-form approximations for the
associated function~$\Lambda(\xi)$ are derived in Section~\ref{sec:find_lambda}, which 
completes the proposed algorithm.
Simulation results as well as the estimated BLER
are compared in Section~\ref{sec:sim}, which reveals the effectiveness
of the RCA-based polar code design using the proposed algorithm.
Finally, concluding remarks are given in Section~\ref{sec:conc}.

\section{Polar Codes and Reciprocal Channel Approximation}
\label{sec:Pol}

\subsection{Polar Codes}

\begin{figure} [tbp]
\begin{center}
\includegraphics[width=.7\textwidth,clip]{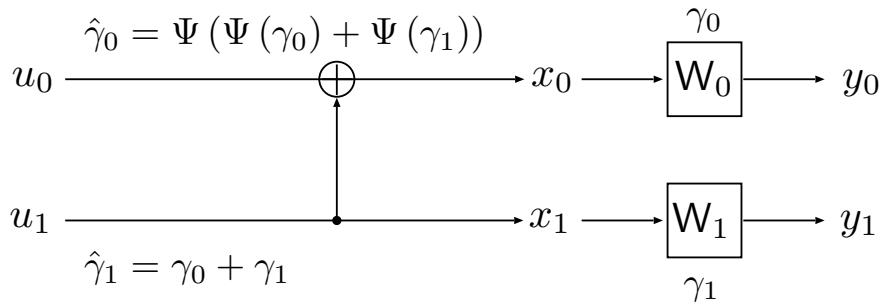}
\end{center}
 \caption{Polar encoder and associated notations for $N=2$.}
\label{fig:enc}
\end{figure}

We start with the simplest binary polar code of codeword (or block)
length $N=2$  
(Ar{\i}kan's kernel) shown in Fig.~\ref{fig:enc},
where the information bits $u_0, u_1 \in {\mathbb F}_2$ 
and the coded bits $x_0, x_1 \in {\mathbb F}_2$ are related by
\begin{align}
\begin{cases}
x_0 & = u_0 + u_1, \\
x_1 & =  u_1 .
\end{cases}
\label{eq:kernel}
\end{align}
The coded bits $x_0$ and $x_1$ are modulated by BPSK 
and are transmitted over AWGN channels, denoted by ${\sf W}_0$ and ${\sf W}_1$, 
where the SNRs of the channels are $\gamma_0$ and $\gamma_1$, respectively,
as illustrated in Fig.~\ref{fig:enc}, with the received symbols given by
$y_0$ and $y_1$.

Assuming that $u_0$ is decoded first, 
since
\begin{align}
u_0 & = x_0 + x_1 ,
\label{eq:kernel2}
\end{align}
$u_0$ can be seen as a check node connected to $x_0$ and $x_1$
in the Tanner graph.
Once the estimate of $u_0$, denoted by
$\hat{u}_0 \in {\mathbb F}_2$, is given,
$u_1$ can be uniquely determined by either of $x_0$ or $x_1$,
i.e., $u_1$ is a variable node connected to both $x_0$ and $x_1$.

Polar codes of length $N=2^n$ for an integer $n>1$ can be obtained by the recursive 
application of the above kernel~\cite{arikan09}.

\subsection{RCA}

Following~\cite{divsalar2009}, we briefly describe the principle of RCA.
Let us first consider a BEC where the erasure probability is $e^{-s}$
and the non-erasure probability $e^{-r}$, i.e., $e^{-s} + e^{-r} = 1$, for $s, r>0$.
Let $C(s)$ denote the capacity of this channel, i.e., $C(s) = 1 -e^{-s}$.
Then, it follows that $C(s) + C(r) = 1$.
Let $s_i$ denote the above parameter $s$ for the bit $x_i$
transmitted over the corresponding BEC
and let us define $r_i$ such that $e^{-s_i} + e^{-r_i} = 1$.
For a variable node $u_1$ connected to $x_0$ and $x_1$,
its erasure probability is given by
$e^{-s_0} e^{-s_1} = e^{-(s_0 + s_1)}$, i.e., it is characterized by the sum of 
$s_0$ and $s_1$.
For a check node $u_0$ connected to $x_0$ and $x_1$,
its corresponding {\em non-erasure} probability is given by
$(1 - e^{-s_0}) (1 - e^{-s_1}) = e^{-(r_0 + r_1)}$, i.e., 
it is also characterized by the sum of $r_0$ and $r_1$.
In other words, for BEC, the parameter $s$ is additive for variable nodes,
whereas the parameter $r$ is additive for check nodes~\cite{divsalar2009}.

We now apply the above RCA concept to the binary-input
AWGN (BI-AWGN) channel. We define $C(\gamma)$ as the capacity of the BI-AWGN channel
with the signal-to-noise power ratio (SNR) given by $\gamma$.
Similar to BEC, 
the additive property holds for the parameter $\gamma$
in the case of the variable node 
as long as all the connecting nodes are
associated with mutually independent AWGN channels. 
This stems from the fact that the LLR is additive at the variable node,
and both mean and variance of the LLR corresponding to BI-AWGN channels
are proportional to the channel SNR~\cite{ochiai2021}.
Let us now define the SNR parameter $\lambda$ corresponding to the check node
such that
$C(\gamma)+ C(\lambda) = 1$, and assume that
the additive property also holds for the parameter $\lambda$ at the check node.
Since this does not strictly hold for general BI-AWGN channels, 
this is an {\em approximation}, which is thus
referred to as the reciprocal channel approximation~(RCA)
in~\cite[Section~7.4]{chung_diss_2000}.

For a given channel SNR $\gamma$, the parameter $\lambda$ can be found by solving
\begin{align}
 C(\lambda) = 1 - C(\gamma),
\label{eq:Cl_g}
\end{align}
for $\lambda$. 
Since $C(\gamma)$ is a strictly increasing
function for $\gamma>0$,
we can define its inverse function $C^{-1}(\cdot)$ and write 
\begin{align}
\lambda = C^{-1}\left( 1 - C(\gamma) \right) \triangleq \Psi(\gamma).
\label{eq:Psi}
\end{align}
The above function corresponds to the {\em reciprocal channel mapping}
introduced in~\cite[Definition~7.3]{chung_diss_2000}, 
also referred to as {\em self-inverting reciprocal energy function} in~\cite{divsalar2009},
applied to BI-AWGN channels.
Our goal is to find a suitable expression for the function $\Psi(\gamma)$ 
that can support a wide range of SNR value $\gamma$ with high accuracy.
Note that since \eqref{eq:Cl_g} holds with $\lambda$ and $\gamma$ interchanged,
one can easily verify that
\begin{align}
\Psi\left(
\Psi\left(\gamma\right)
\right) = \gamma,
\end{align}
i.e., $\Psi(\gamma)$ is a self-inverse function satisfying $\Psi(\gamma) = \Psi^{-1}(\gamma)$~\cite{chung_diss_2000}.

If $\gamma_0$ and $\gamma_1$ are the SNRs of the channels where the bits $x_0$ and $x_1$ are 
transmitted, respectively, then the SNR corresponding to 
the variable node $u_1$, denoted by $\hat{\gamma}_1$, is 
expressed as 
\begin{align}
\hat{\gamma}_1 = \gamma_0 + \gamma_1.
\end{align}
On the other hand,
the SNR corresponding to the check node $u_0$,
denoted by $\hat{\gamma}_0$, is similarly expressed 
based on the RCA principle~\cite[Section~7.4]{chung_diss_2000} as
\begin{align}
 \Psi \left(
\hat{\gamma}_0
\right)
=
\Psi
\left(
\gamma_0
\right)
+
\Psi\left(
\gamma_1
\right),
\end{align}
from which we obtain
\begin{align}
\hat{\gamma}_0 =
 \Psi^{-1} \left(
\Psi\left(
\gamma_0
\right)
+
\Psi\left(
\gamma_1
\right)
\right)
=
 \Psi \left(
\Psi
\left(
\gamma_0
\right)
+
\Psi\left(
\gamma_1
\right)
\right).
\end{align}
The above relationship between the pairs $(\gamma_0, \gamma_1)$ and 
$(\hat{\gamma}_0,\hat{\gamma}_1)$ is illustrated in Fig.~\ref{fig:enc}.

In what follows, we consider the case when $\gamma$ is extremely large or small
due to the polarization effect. 
From the viewpoint of numerical evaluation
and similar to~\cite{ochiai2021},
it would be 
convenient to express $\gamma$ in the log domain.
Therefore, we define $\xi \triangleq \ln \gamma$ and also introduce the function
\begin{align}
 \Lambda(\xi) \triangleq \ln \Psi\left(e^{\xi}\right).
\label{eq:Lambda_def}
\end{align}
For the check node, the output $\hat{\xi}_0$ for a given pair of inputs
$\xi_0 = \ln \gamma_0$ and $\xi_1 = \ln \gamma_1$ 
can be expressed as
\begin{align}
\hat{\xi}_0 & = 
\Lambda
\left(
\max\left(
\Lambda\left(\xi_0\right),
\Lambda\left(\xi_1\right)
\right)
+\ln
\left(
1 + 
e^{-\left|
\Lambda\left(\xi_1\right)
-
\Lambda\left(\xi_0\right)\right|
}
\right)
\right).
\end{align}
On the other hand, for the variable node, the output $\hat{\xi}_1$ is expressed as
\begin{align}
\hat{\xi}_1 
&= \max(\xi_0, \xi_1) + \ln\left( 1 + e^{- \left|\xi_0 - \xi_1\right|}\right) .
\end{align}

\subsection{General Algorithm}
In the case of a binary polar code of length $N=2^n$,
let $\gamma_0,\gamma_1, \ldots, \gamma_{N-1}$
denote the channel SNRs of the coded bits
$x_0, x_1, \ldots, x_{N-1}$.
Then 
the SNRs for the input bits $u_0, u_1, \ldots, u_{N-1}$,
denoted by $\hat{\gamma}_0,\hat{\gamma}_1, \ldots, \hat{\gamma}_{N-1}$,
can be obtained by the well-known recursive procedure in~\cite{arikan09}.
The corresponding RCA algorithm when each channel has a distinct SNR value
is summarized in {\bf Algorithm~\ref{alg:RCA}}.
The algorithm can be significantly simplified when
all the channels have the same SNR (or design SNR),
i.e., $\gamma_0 = \gamma_1 = \cdots = \gamma_{N-1}$,
which is summarized in {\bf Algorithm~\ref{alg:RCA1}}.

The remaining questions are: how can one calculate the function
$\Lambda(\xi)$ accurately with computational efficiency,
i.e., without resorting to numerical integration,
and how well does the algorithm operate
compared to other known 
approaches of similar complexity? These will be addressed in the subsequent
sections.

  \begin{algorithm}[t]
   \caption{Channel Polarization with RCA (for Distinct Channel SNRs).}
   \begin{algorithmic}[1]
    \REQUIRE $n = \log_2 N$, $\xi[0], \xi[1], \ldots, \xi[N-1] $ as 
$\ln \gamma_{0}, \ln \gamma_{1}, \ldots, \ln \gamma_{N-1}$.
    \ENSURE $\xi[0], \xi[1], \ldots, \xi[N-1]$ as $\ln \hat{\gamma}_0, 
 \ln \hat{\gamma}_1,   \ldots, \ln \hat{\gamma}_{N-1}$ 
    \FOR{$i=1$ : $n$}
    \STATE $J \leftarrow 2^{i}$
    \FOR{$k=0$ : $N/J - 1$}
    \FOR{$j=0$ : $J/2 - 1$}
    \STATE $\xi_0 \leftarrow \xi[k J + j]$ 
    \STATE $\xi_1 \leftarrow \xi[k J + j + J/2]$ 
    \STATE $\Lambda_0 \leftarrow \Lambda( \xi_0 )$ 
    \STATE $\Lambda_1 \leftarrow \Lambda( \xi_1 )$ 
    \STATE $\xi[k J + j] \leftarrow \Lambda\left(\max(\Lambda_0, \Lambda_1) + \ln\left( 1 + e^{- \left|\Lambda_0 - \Lambda_1\right|}\right) \right)$
    \STATE $\xi[k J + j + J/2] \leftarrow  
\max(\xi_0, \xi_1) + \ln\left( 1 + e^{- \left|\xi_0 - \xi_1\right|}\right)$
    \ENDFOR
    \ENDFOR
    \ENDFOR
    \RETURN $\xi[0], \xi[1], \ldots, \xi[N-1]$ 
   \end{algorithmic}
   \label{alg:RCA}
  \end{algorithm}

  \begin{algorithm}[t]
   \caption{Channel Polarization with RCA (for Uniform Channel SNR).}
   \begin{algorithmic}[1]
    \REQUIRE $n = \log_2 N$, $\xi[0] = \ln \gamma_0$
    \ENSURE $\xi[0], \xi[1], \ldots, \xi[N-1]$ as $\ln \hat{\gamma}_0, 
 \ln \hat{\gamma}_1,   \ldots, \ln \hat{\gamma}_{N-1}$ 
    \FOR{$i=1$ : $n$}
    \STATE $J \leftarrow 2^{i}$
    \FOR{$j=0$ : $J/2-1$}
    \STATE $\xi_0 \leftarrow \xi[j]$ 
    \STATE $\Lambda_0  \leftarrow \Lambda( \xi_0 )$ 
    \STATE $\xi[j] \leftarrow \Lambda\left(\Lambda_0 + \ln 2 \right)$ 
    \STATE $\xi[j+J/2] \leftarrow \xi_0 + \ln 2 $
    \ENDFOR
    \ENDFOR
    \RETURN $\xi[0], \xi[1], \ldots, \xi[N-1]$ 
   \end{algorithmic}
   \label{alg:RCA1}
  \end{algorithm}

\section{Capacity Expression}
\label{sec:cap_expression}

\subsection{Approximation for BI-AWGN Capacity}

We start with the BI-AWGN capacity formula. 
Let $E_s$ denote the symbol energy of BPSK and $N_0$ denote the variance
of complex Gaussian noise. 
Let us define the SNR parameter $\gamma$ as 
$\gamma \triangleq E_s/N_0$ in what follows.
To compute the mutual information $C(\gamma)$
between a BPSK input and its corresponding output
over an AWGN channel,
the following equivalent 
$J$-function 
(see~\cite{ten_brink2004})
is often adopted:
\begin{align}
 J(x) &=  1 - \frac{1}{\sqrt{2\pi x^2}} \int_{-\infty}^{\infty}
e^{-\frac{
\left(
t - \frac{x^2}{2}\right)^2
}{2 x^2}} \log_2 \left( 1 + e^{-t}\right) dt,
\label{eq:Jx_def}
\end{align}
where $x$ corresponds to the standard deviation of the LLR
and thus is related to 
$\gamma$ by $x=\sqrt{8\gamma}$.
Then $C(\gamma) =  J\left(\sqrt{8\gamma} \right).$

According to~\cite{ten_brink2004}, the function $J(x)$ is 
well approximated numerically by
\begin{align}
 J(x) &\approx
\begin{cases}
 a_1 x^3 + b_1 x^2 + c_1 x, & 0 \leq x \leq 1.6363, \\
 1 - e^{ a_2 x^3 + b_2 x^2 + c_2 x + d}, & 1.6363 < x \leq 10, \\
1, & x > 10, 
\end{cases}
\label{eq:tenBrink}
\end{align}
with
\begin{align}
a_1 & = -0.0421061,\,\, b_1 = 0.209252,\,\, c_1=-0.00640081,
\nonumber
\\
a_2 & = 0.00181491, \,\, b_2 = -0.142675, \,\, c_2 = -0.0822054,
\,\, d = 0.0549608.
\nonumber
\end{align}
On the other hand, in~\cite{brannstrom2005}, the 
approximation 
\begin{align}
 J (x) & \approx \left( 1 - 2^{-H_1 x^{2 H_2} }\right)^{H_3} ,
\label{eq:Jx_Bran}
\end{align}
was proposed with $H_1=0.3073$, $H_2=0.8935$, and $H_3=1.1064$.

\subsection{New Approximation Formula}

From~\eqref{eq:Jx_def} and $C(\gamma) =  J\left(\sqrt{8\gamma} \right)$,
we define
\begin{align}
U(\gamma) & \triangleq 1 - C(\gamma)  \nonumber \\
 &=
 \frac{1}{4\sqrt{\pi \gamma}}
\int_{-\infty}^{\infty}
e^{ - 
\frac{1}{16\gamma}
\left(
t  - 4\gamma 
\right)^2}
\log_2 
\left(
1 + e^{- t}
\right)
dt.
\label{eq:Lgamma}
\end{align}
In what follows, we divide the range of $\gamma$, ${\cal R} \triangleq (0, \infty)$,
into the four sub-regions ${\cal R}_1, {\cal R}_2, {\cal R}_3, {\cal R}_4$,
where ${\cal R}_1 = (0,\Gamma_1)$, ${\cal R}_2 = [\Gamma_1, \Gamma_2)$,
${\cal R}_3 = [\Gamma_2, \Gamma_3)$, and
${\cal R}_4 = [\Gamma_3, \infty)$, with $\Gamma_1, \Gamma_2, \Gamma_3$ 
representing 
appropriate boundaries to be determined numerically.

\subsubsection{Low SNR Case}

For the sub-region ${\cal R}_1$, i.e., when $\gamma \approx 0$, 
by Maclaurin series
expansion we obtain 
\begin{align}
C(\gamma) &\approx
\frac{1}{\ln 2}
\left(
\gamma - \gamma^2 + \frac{4}{3} \gamma^3
- \frac{10}{3} \gamma^4 + \frac{208}{15}\gamma^5 
\right) .
\end{align}
By truncating up to the third order, 
we may approximate $U(\gamma)$ as
\begin{align}
U_1(\gamma)  \triangleq 
1 - 
\frac{1}{\ln 2}
\left(
\gamma - \gamma^2 + \frac{4}{3} \gamma^3
\right), \quad \gamma < \Gamma_1.
\label{eq:L1}
\end{align}
By setting the upper boundary as $\Gamma_1 = 0.04$,
the value of $U(\gamma)$ at $\gamma = \Gamma_1$ is calculated by numerical
integration as
\begin{align}
U(\Gamma_1)   \approx 
0.9444880,
\end{align}
whereas the corresponding approximation according to \eqref{eq:L1}
is given by
\begin{align}
{\sf U}_1 \triangleq U_1(\Gamma_1) \approx 0.9444774 .
\end{align}
Therefore, the approximation error
of $U(\gamma)$ by $U_1(\gamma)$ in the region ${\cal R}_1$
is bounded by
\begin{align}
\left|
U(\gamma)  - U_1(\gamma)
\right|
< 1.1 \times 10^{-5} , \quad \gamma < \Gamma_1 .
\label{eq:approx_cond}
\end{align}

\subsubsection{High SNR Case}

We next consider the sub-region ${\cal R}_4$, i.e., when $\gamma$ becomes larger than
some boundary $\Gamma_3$.
By applying the series expansion of the exponential function
$e^x = \sum_{k=0}^{\infty} \frac{x^k}{k!}$ to $U(\gamma)$ of \eqref{eq:Lgamma},
we obtain
\begin{align}
U(\gamma) &= 
 \frac{e^{-\gamma}}{4\sqrt{\pi \gamma}}
\sum_{k=0}^{\infty} 
\frac{(-1)^k}{k!\left(16\gamma\right)^k}
c_k \nonumber\\
& =
 \frac{e^{-\gamma}}{4\sqrt{\pi \gamma}}
\left\{
c_0 - 
\frac{c_1}{16 \gamma} + 
\frac{c_2}{2\cdot\left(
16 \gamma\right)^2} 
- \cdots
\right\},
\label{eq:Lgamma_high}
\end{align}
where
\begin{align}
c_k \triangleq \int_{-\infty}^{\infty}
e^{
\frac{t}{2}}
t^{2k}
\log_2 
\left(
1 + e^{- t}
\right)
dt.
\end{align}
The above expression agrees with~\cite[Problem~4.12]{richardson_modern_2008}.
Note that $c_k$ for specific values of $k$ can be expressed in closed form,
e.g.,
\begin{align}
 c_0 = \frac{2\pi}{\ln 2}, \quad
c_1 = \frac{2 \pi \left(8 + \pi^2\right)}{\ln 2 }, \quad
c_2 = \frac{2 \pi \left(384 + 48 \pi^2 + 5 \pi^4\right)}{\ln 2 }.
\end{align}
Although the approximation becomes tighter as we incorporate 
more terms in \eqref{eq:Lgamma_high},
it becomes infeasible to find the corresponding inverse expression in closed form.
Therefore, we adopt the expression
\begin{align}
U_4 (\gamma) \triangleq
\alpha
 \frac{e^{-\gamma}}{\sqrt{\gamma}},
\quad \gamma > \Gamma_3,
\end{align}
where $\alpha$ and $\Gamma_3$ are appropriate constants to be determined.
By setting the boundary as $\Gamma_3 = 10$,
the value at the boundary becomes
\begin{align}
{\sf U}_3 \triangleq
U(\Gamma_3) \approx 1.667 \times 10^{-5},
\end{align}
from which we fix the constant $\alpha$ as
\begin{align}
\alpha \approx 1.16125142,
\label{alpha_approx}
\end{align}
such that $U(\Gamma_3) \approx U_4(\Gamma_3)$.
Note that this is different from the exact coefficient of the first term in 
\eqref{eq:Lgamma_high}, which is given by
\begin{align}
\alpha_0 \triangleq c_0/(4 \sqrt{\pi}) 
=\frac{\sqrt{\pi} }{2 (\ln 2) }
\approx 1.27856,
\label{c0_asympto}
\end{align}
and thus the exact asymptotic form for large $\gamma$ would be
\begin{align}
U(\gamma) \to 
\frac{\sqrt{\pi} }{2 (\ln 2) }
 \frac{e^{-\gamma}}{\sqrt{\gamma}},
\end{align}
as discussed in~\cite{richardson_modern_2008,lozano2006}. 
As $\gamma$ increases, the use of $\alpha_0$ may become more accurate eventually,
but for our purpose, the use of 
\eqref{alpha_approx} may be more suitable as the piece-wise continuity 
of the approximate function can be guaranteed.

\subsubsection{Moderate SNR Case}

We first note that the main advantage of the capacity approximation form~\eqref{eq:Jx_Bran}
is that its inverse function can also be 
expressed in closed form.
We thus adopt this form for the remaining sub-regions ${\cal R}_2$ and ${\cal R}_3$
and define the functions for the two adjacent regions as
\begin{align}
 U_2(\gamma) &\triangleq
1 - \left(
1 - e^{- H_{2,1} \gamma^{H_{2,2}}}
\right)^{H_{2,3}}
 , \quad  \Gamma_1 < \gamma < \Gamma_2, \\
U_{3}(\gamma)  
&\triangleq
1 - \left(
1 - e^{- H_{3,1} \gamma^{H_{3,2} }}
\right)^{H_{3,3}}
, \quad  \Gamma_2 < \gamma < \Gamma_3,
\end{align}
where the constants $H_{i,j}$ for $i\in\{2,3\}$ and $j\in\{1,2,3\}$
should be determined appropriately
depending on the boundary $\Gamma_2$.
In what follows, we fix the boundary
as $\Gamma_2 = 1$ for simplicity of numerical evaluation.
Then the precise value of $U(\gamma)$ at the boundary is obtained by numerical integration
as
\begin{align}
{\sf  U}_2 \triangleq
U(\Gamma_2)
\approx 0.2785484.
\end{align}
Furthermore, in order to guarantee the piece-wise continuity of the 
approximate function for $U(\gamma)$, it is required that
\begin{align}
U_2(\Gamma_1) & = {\sf U}_1, \quad U_3(\Gamma_3)  = {\sf U}_3 .
\label{eq:U2_U3}
\end{align}
Let us first consider the function $U_{2}(\gamma)$.
By the relationships at the boundaries,  one may express
\begin{align}
H_{2,1} &=  - 
\ln\left(
1 - 
\left(
1 - {\sf U}_2
\right)^{\frac{1}{H_{2,3}}}
\right),
\label{eq:H1d}\\
H_{2,2}
& =
\left[
\ln 
\left(
\frac{\Gamma_2}{\Gamma_1}
\right)
\right]^{-1}
\ln
\left\{
 \frac{
\ln\left(
1 - 
\left(
1 - {\sf U}_2
\right)^{\frac{1}{H_{2,3}}}
\right)
}{
\ln\left(
1 - 
\left(
1 - {\sf U}_1
\right)^{\frac{1}{H_{2,3}}}
\right)
}\right\}.
\label{eq:H2d}
\end{align}
As a consequence, $U_2(\gamma)$ can be expressed as a function of
$\gamma$ and $H_{2,3}$, which we explicitly write as
$U_2(\gamma;H_{2,3})$.
Based on the minimization of $\infty$-norm, we may optimize the coefficients $H_{2,3}$ as
\begin{align}
{H}_{2,3} = \arg \min_{H} \max_{\Gamma_1 < \gamma < \Gamma_2} \left|
U_2(\gamma;{H}) - U(\gamma)
\right|,
\label{eq:optimize}
\end{align}
based on which
we numerically obtain
\begin{align}
H_{2,1} = 1.396634, \, H_{2,2} = 0.872764, \, H_{2,3} = 1.148562.
\nonumber
\end{align}
Likewise,
by denoting $U_3(\gamma)$ as a function of $H_{3,3}$,
i.e., $U_3(\gamma;{H}_{3,3})$,
since
\begin{align}
H_{3,1} &=  - 
\ln\left(
1 - 
\left(
1 - {\sf U}_2
\right)^{\frac{1}{H_{3,3}}}
\right),
\\
H_{3,2}
&=
\left[
\ln 
\left(
\frac{\Gamma_3}{\Gamma_2}
\right)
\right]^{-1}
\ln
\left\{
 \frac{
\ln\left(
1 - 
\left(
1 - {\sf U}_3
\right)^{\frac{1}{H_{3,3}}}
\right)
}{
\ln\left(
1 - 
\left(
1 - {\sf U}_2
\right)^{\frac{1}{H_{3,3}}}
\right)
}\right\},
\label{eq:H2dd}
\end{align}
one may determine $H_{3,3}$ according to
\begin{align}
{H}_{3,3} = \arg \min_{H} \max_{\Gamma_2 < \gamma < \Gamma_3} \left|
U_3(\gamma;{H}) - U(\gamma)
\right|,
\label{eq:optimize2}
\end{align}
and thus numerically obtain
\begin{align}
H_{3,1} = 1.266967, \quad H_{3,2} = 0.938175, \quad H_{3,3} = 0.986830.
\nonumber
\end{align}

\subsubsection{Summary}

We summarize the closed-form 
piece-wise continuous approximate function $\hat{U}(\gamma)$
for $U(\gamma)$: 
\begin{align}
\hat{U}(\gamma)
&=
\begin{cases}
U_1(\gamma)
= 1 - 
\frac{1}{\ln 2}
\left(
\gamma - \gamma^2 + \frac{4}{3} \gamma^3
\right), & \gamma < \Gamma_1, \\
U_2(\gamma) =
1 - \left(
1 - e^{- H_{2,1}\gamma^{H_{2,2} }}
\right)^{H_{2,3}}, & \Gamma_1  \leq  \gamma < \Gamma_2,  \\
U_3(\gamma) =
1 - \left(
1 - e^{- H_{3,1} \gamma^{H_{3,2}}}
\right)^{H_{3,3}}, & \Gamma_2  \leq  \gamma < \Gamma_3, \\
U_4(\gamma) =
\alpha
\frac{e^{-\gamma}}{\sqrt{\gamma}},
 &   \gamma \geq \Gamma_3,
\end{cases}
\label{eq:L_final}
\end{align}
where $\Gamma_1 = 0.04, \Gamma_2 = 1, \Gamma_3 = 10$.
As a consequence, the approximate formula of the capacity for the BI-AWGN channel,
$\hat{C}(\gamma) \triangleq 1 - \hat{U}(\gamma)$, can be summarized as
\begin{align}
\hat{C}(\gamma)
&=
\begin{cases}
\frac{1}{\ln 2}
\left(
\gamma - \gamma^2 + \frac{4}{3} \gamma^3
\right), & 
\gamma \in {\cal R}_1,
\\
 \left(
1 - e^{- H_{2,1}\gamma^{H_{2,2} }}
\right)^{H_{2,3}}, & 
\gamma \in {\cal R}_2,
\\
 \left(
1 - e^{- H_{3,1} \gamma^{H_{3,2}}}
\right)^{H_{3,3}}, & 
\gamma \in {\cal R}_3,
\\
1- \alpha
\frac{e^{-\gamma}}{\sqrt{\gamma}},
& 
\gamma \in {\cal R}_4.
\end{cases}
\label{eq:C_final}
\end{align}

\subsection{Numerical Results}

\begin{figure} [tbp]
\begin{center}
\includegraphics[width=.85\textwidth,clip]{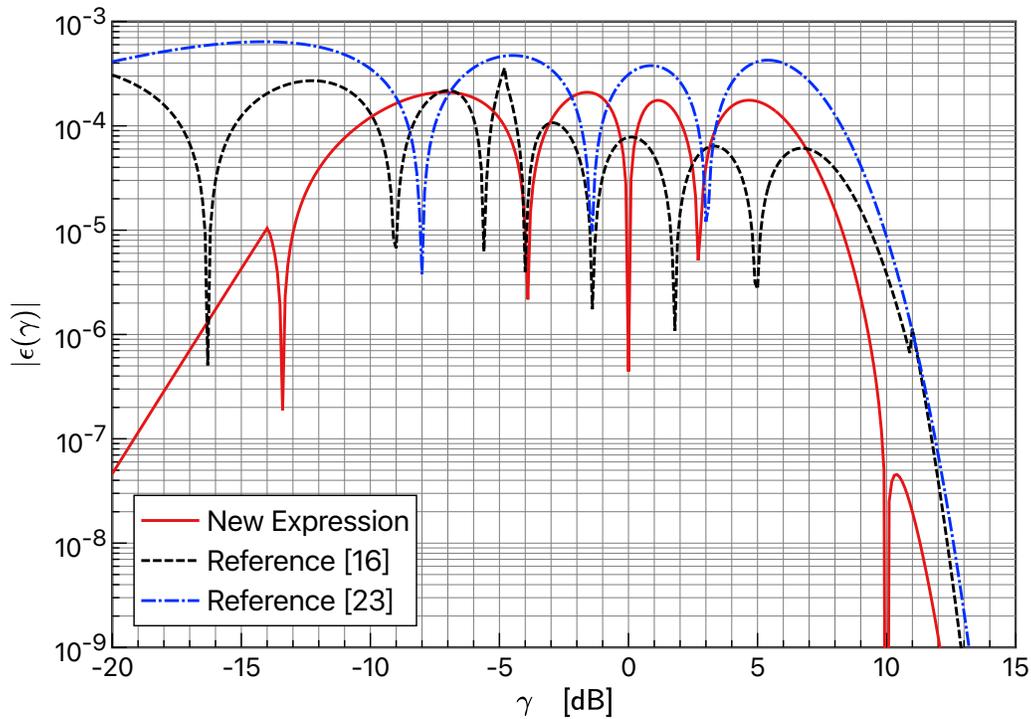}
\end{center}
 \caption{Approximation error of the proposed expression. 
Those based on ten Brink {\em et al.}~\cite{ten_brink2004} and
Br\"annstr\"om {\em et al.}~\cite{brannstrom2005} are also shown for comparison.}
\label{fig:error}
\end{figure}

To investigate the accuracy of the developed approximation,
we define the error function as
\begin{align}
 \epsilon(\gamma) \triangleq U(\gamma) - \hat{U}(\gamma)
=  \hat{C}(\gamma) - C(\gamma) .
\end{align}
We calculate $C(\gamma)$ based on the numerical integration and 
compare with the developed expression as well as 
\eqref{eq:tenBrink} of~\cite{ten_brink2004} and
\eqref{eq:Jx_Bran} of~\cite{brannstrom2005}.
In Fig.~\ref{fig:error}, we plot the absolute value
of the error function $\epsilon(\gamma)$ with respect to $\gamma$ in dB.
We observe that the absolute values of the approximation
error for all the three expressions are
less than $10^{-3}$, and monotonically decrease
for higher and lower SNR regions. Nevertheless, the most significant difference
of the proposed expression is that the error becomes 
much less than $10^{-5}$ as SNR decreases
since we selected the associated parameters such that
the condition~\eqref{eq:approx_cond} holds.
Also, since the proposed expression $\hat{C}(\gamma)$ in \eqref{eq:C_final}
is designed to be piece-wise continuous with respect to $\gamma$,
we also observe that the error function is piece-wise continuous as well.
By comparison, as observed in Fig.~\ref{fig:error},
the approximation
based on~\eqref{eq:tenBrink}
(introduced in~\cite{ten_brink2004})
exhibits discontinuities at
the boundaries corresponding to $x=1.6363$
and $x=10$, i.e., $\gamma = -4.7536$\,dB
and $\gamma =10.969$\,dB, respectively.

\section{Derivation of Closed-Form Expression for $\Lambda(\xi)$}
\label{sec:find_lambda}

Our next step is to find a suitable approximate expression for $\Lambda(\xi)$
introduced in~\eqref{eq:Lambda_def}.
We first consider an approximation for its constituent function  $\Psi(\gamma)$.
From \eqref{eq:Psi}, we have
\begin{align}
{\Psi}(\gamma)
=
C^{-1}\left(
1 -  C(\gamma) 
\right)
=
C^{-1}\left(
U(\gamma) 
\right).
\end{align}
Therefore, it is necessary to find the inverse function for
$\hat{C}(\gamma)$ defined in \eqref{eq:C_final}.

Let us first consider the case $\gamma \in {\cal R}_1$.
From \eqref{eq:C_final},
the inverse function for 
the equation 
$\hat{C}(\gamma) = c$
can be expressed as
\begin{align}
\hat{C}^{-1}(c)  & =  
\frac{1}{4}
\left(
1 - \frac{3}{A(c)} +A(c)
\right), \quad c < {\sf C}_1,
\end{align}
where ${\sf C}_1 \triangleq 1 - {\sf U}_1$ and
\begin{align}
 A(c)& \triangleq
\left(
-5 + 24 \left(\ln 2\right)
c + 2 \sqrt{13 + 12 \left(\ln 2\right)
c (12 \left(\ln 2\right)
c - 5)} \right)^{\frac{1}{3}} .
\label{eq:AI}
\end{align}

On the other hand, when $\gamma \in {\cal R}_2$, we have
\begin{align}
\hat{C}^{-1}(c)  & =  
\left[
-\frac{1}{H_{2,1}}
 \ln
\left(
1 - 
I^{\frac{1}{H_{2,3}}}
\right)
\right]^{\frac{1}{H_{2,2}}}, \quad {\sf C}_1 < c < {\sf C}_2,
\end{align}
where ${\sf C}_2 = 1 - {\sf U}_2$.
The function $\hat{C}^{-1}(c)$
in the case of $\gamma \in {\cal R}_3$ can be derived in a similar manner.
Finally, in the case of $\gamma \in {\cal R}_4$, we have
\begin{align}
\hat{C}^{-1}(c)  & =  
\frac{1}{2} W_0\left(
2 \left(
\frac{\alpha}{1 - c}
\right)^2
\right),\quad c > {\sf C}_3,
\label{eq:c_W0}
\end{align}
where ${\sf C}_3 = 1 - {\sf U}_3$ and
$W_0(x)$ is 
the corresponding Lambert W function~\cite{corless1996}, i.e.,
the value of $w$ that satisfies $w e^w = x$ for a given $x>0$.

In summary, we have
\begin{align}
\hat{C}^{-1}(c) 
= \begin{cases}
\frac{1}{4}
\left(
1 - \frac{3}{A(c)} +A(c)
\right),
& c < {\sf C}_1  \approx 0.055523, 
\\
\left[
-\frac{1}{H_{2,1}}
 \ln
\left(
1 - 
c^{\frac{1}{H_{2,3}}}
\right)
\right]^{\frac{1}{H_{2,2}}},
& {\sf C}_1 \leq c < {\sf C}_2 \approx 0.721452,\\
\left[
-\frac{1}{H_{3,1}}
 \ln
\left(
1 - 
c^{\frac{1}{H_{3,3}}}
\right)
\right]^{\frac{1}{H_{3,2}}},
& {\sf C}_2 \leq c <
{\sf C}_3  \approx 0.999983,\\
\frac{1}{2} W_0\left(
2 \left(
\frac{\alpha}{1 - c}
\right)^2
\right), & 
c \geq {\sf C}_3.
  \end{cases}
\label{eq:C_inv}
\end{align}

By substituting $\hat{U}(\gamma)$ of \eqref{eq:L_final}
into $c$ of \eqref{eq:C_inv}, we obtain
the approximate value of
$\Psi(\gamma) = C^{-1}(U(\gamma))$ for a given $\gamma$,
which we denote by $\hat{\Psi}(\gamma) \triangleq \hat{C}^{-1}(\hat{U}(\gamma))$.

In what follows, we derive 
simple asymptotic closed-form expressions of
$\hat{\Psi}(\gamma)$ and thus $\Lambda(\xi)$
when $\gamma$ becomes extremely small or large in order to simplify the computation
with negligible loss of accuracy.
Also, as we will see, it is not actually necessary to implement the 
$W_0(\cdot)$ function 
for polar code construction.

\subsection{For Small $\gamma$}

We consider the case where $\gamma$ is small 
and satisfies $\gamma \leq \Gamma_0$ for some boundary
$\Gamma_0$ such that 
the condition $c=\hat{U}(\gamma) \geq {\sf C}_3$ holds
in \eqref{eq:C_inv}.
Then, it follows from \eqref{eq:L_final}
that
\begin{align}
\hat{\Psi}(\gamma) = 
\frac{1}{2} W_0\left(
2 \left(
\frac{\alpha}{
\frac{1}{\ln 2}
\left(
\gamma - \gamma^2 + \frac{4}{3} \gamma^3
\right)
}
\right)^2
\right), \quad \gamma < \min\left(
\Gamma_{0},\Gamma_1 \right),
\label{eq:Psi_LOW}
\end{align}
where $\Gamma_0$ corresponds to the value of $\gamma$
that satisfies
$\hat{U}(\gamma)={\sf C}_3$.
By numerically solving the equation 
\begin{align}
\frac{1}{\ln 2}
\left(
\Gamma_0 - \Gamma_0^2 + \frac{4}{3} \Gamma_0^3
\right)  = {\sf U}_3,
\end{align}
with respect to $\Gamma_0$, we obtain
\begin{align}
\Gamma_{0} \approx  1.21974 \times 10^{-5},
\end{align}
and thus we observe that $\Gamma_0 \ll \Gamma_1$.
By defining $f(\gamma) \triangleq \gamma - \gamma^2 + \frac{4}{3}\gamma^3$,
we may express \eqref{eq:Psi_LOW} as
\begin{align}
\hat{\Psi}(\gamma) = 
\frac{1}{2}W_0 \left(
2 \alpha^2
\left(
\ln 2
\right)
^2
 f^{-2}(\gamma)
\right) , \quad \gamma < \Gamma_0.
\label{eq:W0}
\end{align}
Since the argument of the function $W_0(\cdot)$ in \eqref{eq:W0} becomes larger
as $\gamma$ decreases, we may invoke the following asymptotic
form of $W_0(x)$ for large $x$~\cite{corless1996}:
\begin{align}
 W_0(x) \approx \ln x - \ln \ln x + \frac{\ln \ln x}{\ln x} +
\cdots.
\label{eq:W0asympto}
\end{align}
On applying~\eqref{eq:W0asympto} to~\eqref{eq:W0},
we note that
\begin{align}
\ln 
\left(
2 
\alpha^2
\left(
 \ln 2\right)^2
 f^{-2}(\gamma)
\right)
=
\ln 2 
+ 2 \ln 
(\ln 2)
+ 2 \ln \alpha 
 - 2 \ln \gamma
- 2 \ln
\left(
1 - \gamma + \frac{4}{3} \gamma^2
\right),
\end{align}
where the last term in the right hand side turns out to be negligible
in the range of $\gamma < \Gamma_0$. Thus, by
defining the dominant terms as 
\begin{align}
B(\xi) \triangleq
\ln 2 + 2 
\left(
\ln 
\alpha
+
\ln\left(
 \ln 2\right)
\right)
- 2 \xi,
\end{align}
where $\xi \triangleq \ln \gamma$,
we may equivalently express $\ln \hat{\Psi}(\gamma)$ with respect
to $\xi$, denoted by $\Lambda(\xi)$, as
\begin{align}
\Lambda(\xi)
\approx 
\ln
\left(
B(\xi) 
+
\left(
\frac{1}{B(\xi)}
-1
\right)
\ln B(\xi) 
\right)
- \ln 2.
\label{eq:Psi_low}
\end{align}
The above approximation is valid in the range of $\gamma < \Gamma_0$, i.e.,
\begin{align}
\xi
<
\Xi_0 \triangleq
\ln \Gamma_{0} \approx
 -11.3143 .
\end{align}

\subsection{For Large $\gamma$}

When $\gamma \geq \Gamma_3$, i.e., $\gamma \in {\cal R}_4$, 
we observe from~\eqref{eq:L_final} with~\eqref{eq:U2_U3}
that $\hat{U}(\gamma) \leq {\sf U}_3$ and thus 
\begin{align}
\hat{\Psi}(\gamma) = 
\frac{1}{4}
\left(
1 - \frac{3}{A(c)} +A(c)
\right), \quad
c = 
\alpha \frac{e^{-\gamma}}{\sqrt{\gamma}}.
\label{eq:Psi_high}
\end{align}
For $A(c)$ in \eqref{eq:AI} with small $c$, since
\begin{align}
\sqrt{13 + 12 
\left(
\ln 2 \right)
c \left(12 
\left(
\ln 2
\right)
 c - 5\right)} 
\approx \sqrt{13} - \frac{30 }{\sqrt{13}} \left(\ln 2\right)c, 
\end{align}
we have
\begin{align}
 A(c)& \approx
\left(
-5 + 2 \sqrt{13}
\right)^{\frac{1}{3}}
\left(
1 + \frac{4}{\sqrt{13}} \left(\ln 2\right) c
\right),
\label{eq:Ac_approx}
\end{align}
and by substituting \eqref{eq:Ac_approx}
into \eqref{eq:Psi_high}, we obtain
\begin{align}
\hat{\Psi}(\gamma)
& \approx 
\left(\ln 2\right)
 c \left(
1 - 2 \frac{\sqrt{13} + 1}{13 + 4 \sqrt{13} 
\left(\ln 2\right)
c} 
\left(\ln 2\right)
 c
\right).
\label{eq:Psi_2}
\end{align}
Taking the logarithm of both sides, we have
\begin{align}
\ln \hat{\Psi}(\gamma)
&
\approx 
\ln
 c
+
\ln 
\left(\ln 2\right)
+ \ln \left(1
 - 2 \frac{\sqrt{13} + 1}{13 + 4 \sqrt{13} \left(\ln 2\right) c} \left(\ln 2\right) c 
\right)
.
\label{eq:ln_Phi}
\end{align}
By substituting 
$c = 
\alpha \frac{e^{-\gamma}}{\sqrt{\gamma}}$ and 
noticing that the third term in the right-hand side of \eqref{eq:ln_Phi}
becomes negligible as $\gamma$ increases,
we have
\begin{align}
\ln \hat{\Psi}(\gamma)
&
\approx 
 \ln \alpha 
+
 \ln (\ln 2) 
- \gamma - \frac{1}{2}\ln \gamma,
\label{eq:lnPhi}
\end{align}
or in terms of $\xi$ as
\begin{align}
\Lambda(\xi) 
\approx
 \ln \alpha 
+
 \ln (\ln 2) 
- e^{\xi} - \frac{1}{2}\xi.
\label{eq:L_large}
\end{align}
Note that \eqref{eq:lnPhi} indicates that $\hat{\Psi}(\gamma)$ becomes infinitesimal
as $\gamma$ increases.

\subsection{Complete RCA Algorithm}

In the previous section, the RCA algorithm 
was summarized as {\bf Algorithm~\ref{alg:RCA}} and
{\bf Algorithm~\ref{alg:RCA1}}, depending 
on the condition of the channel SNR.
Based on the mathematical derivations in this section,
the calculation process of the function $\Lambda(\xi)$ 
is summarized as {\bf Algorithm~\ref{alg:findPhi}}.
Note that the complexity of GA-based approaches generally involves
the use of an inverse function that cannot be described (or accurately approximated) 
using a closed form expression, and thus should be performed by resorting to
some iterative algorithms (such as the bisection method). On the other hand,
the proposed approach does not require such numerical algorithms and thus 
computational effort is in general lower than those based on GA (or IGA).

\begin{algorithm}[tbp]
   \caption{Calculating $\Lambda(\xi)$}
   \begin{algorithmic}[1]
    \REQUIRE $\xi = \ln \gamma$ 
    \ENSURE $\Lambda(\xi) =\ln \Psi(\gamma)$
    \STATE $\alpha = 1.16125,\, \Gamma_1 = 0.04, \, \Gamma_2 = 1, \, \Gamma_3 = 10, \, \Xi_0  =  -11.3143$
    \STATE ${\sf C}_1  = 0.055523, \, {\sf C}_2 = 0.721452$
    \STATE $H_{2,1} = 1.396634,\, H_{2,2} = 0.872764, \,H_{2,3} = 1.148562 $
    \STATE $H_{3,1} = 1.266967,\, H_{3,2} = 0.938175,\, H_{3,3} = 0.986830 $
    \IF {$\xi < \Xi_0$}
    \STATE $B = \ln 2 + 2 \ln (\ln 2) + 2 \ln \alpha 
  - 2 \xi$
    \RETURN $\ln\left(B +\left(
\frac{1}{B} - 1
\right) \ln B  \right)  - \ln 2$
    \ENDIF
    \STATE $\gamma \leftarrow \exp( \xi )$
    \IF {$\gamma > \Gamma_3$}
    \RETURN $\ln \left(\ln 2\right) + \ln \alpha
 - \gamma - \xi /2$
    \ELSIF {$\gamma < \Gamma_1$}
    \STATE $ U \leftarrow 1 - \frac{1}{\ln 2}( \gamma - \gamma^2 + \frac{4}{3}\gamma^3 )$
    \ELSIF {$\gamma < \Gamma_2$}
    \STATE $ U \leftarrow 1 - \left( 1 - e^{-H_{2,1} \gamma^{H_{2,2}}}\right)^{H_{2,3}}$
    \ELSE
    \STATE $ U \leftarrow 1 - \left( 1 - e^{-H_{3,1} \gamma^{H_{3,2}}}\right)^{H_{3,3}}$
    \ENDIF
    \IF {$U < {\sf C}_1$}
    \STATE $ A =
\left(
-5 + 24 (\ln 2)
U + 2 \sqrt{13 + 12 (\ln 2)
U (12 (\ln 2)
U - 5)} \right)^{\frac{1}{3}}$
    \RETURN $\ln \left(1 - \frac{3}{A} + A\right) - 2 \ln 2$
    \ELSIF {$U < {\sf C}_2$}
    \RETURN ${\frac{1}{H_{2,2}}}
\left[
\ln \left(- 
\ln
\left(
 1 - U^{\frac{1}{H_{2,3}}}\right)
\right)
- \ln H_{2,1}
\right] 
$
    \ELSE
    \RETURN ${\frac{1}{H_{3,2}}}
\left[
\ln \left(- 
\ln
\left(
1 - U^{\frac{1}{H_{3,3}}}\right)
\right)
- \ln H_{3,1}
\right] 
$
    \ENDIF
   \end{algorithmic}
   \label{alg:findPhi}
  \end{algorithm}

\section{Simulation Results}
\label{sec:sim}

In this section, we focus on the AWGN channels with BPSK, 
where the SNRs of all the channels are identical and given by $\gamma_0=E_s/N_0$.
We estimate the SNRs of the polarized channels 
$\hat{\gamma}_0, \hat{\gamma}_1, \ldots, \hat{\gamma}_{N-1}$
by applying {\bf Algorithm~\ref{alg:RCA1}} in Section~\ref{sec:Pol}.
Let us sort $\hat{\gamma}_k$ such that 
\begin{align}
\hat{\gamma}_{I_0(\gamma_0)} \geq \hat{\gamma}_{I_1(\gamma_0)} \geq \cdots \geq 
\hat{\gamma}_{I_{N-1}(\gamma_0)}, 
\end{align}
where the subscript 
$I_k(\gamma_0)$ corresponds to the index of the input bit having the $k$th highest 
SNR (with $k=0$ representing the maximum), with emphasis on the fact
that the channel SNR is $\gamma_0$.
Let us define the index set ${\cal I}_{K}(\gamma_0)$ consisting of the $K$ elements with
highest SNRs as
\begin{align}
{\cal I}_{K}(\gamma_0) 
\triangleq \{ I_0(\gamma_0), I_1(\gamma_0), \ldots, I_{K-1}(\gamma_0)\},
\end{align}
where $R\triangleq K/N$ corresponds to the code rate. 

Similar to the case of GA, we assume that the distribution of the LLR
corresponding to the $k$th input bit 
is modeled as Gaussian with SNR $\hat{\gamma}_k$.
Then, under the assumption that 
all the previous bits in the SC decoding 
are correctly decoded,
the error rate of the $k$th bit can be estimated as
\begin{align}
P_{b, k} =
 Q\left(
\sqrt{
\frac{\hat{\gamma}_k}{2}}\right),
\label{eq:BER_bit_channel}
\end{align}
with the $Q$-function defined as
$Q(x) = 
\frac{1}{2}{\rm{erfc}}\left(
\frac{x}{\sqrt{2}}
\right)=
\frac{1}{\sqrt{2\pi}} \int_{x}^{\infty} e^{-\frac{t^2}{2}} dt$.
The resulting {\em minimum estimated BLER}
can be defined as~\cite{ochiai2021}
\begin{align}
P_{\rm BL} (K, \gamma_0 ) & \triangleq
1 -
\prod_{k \in {\cal I}_{K} (\gamma_0)}
\left(
1-
 Q\left(
\sqrt{
 \frac{\hat{\gamma}_k
}{2}}
\right)
\right)
.
\label{eq:BLER}
\end{align}
Note that the above expression is a function of ${\cal I}_{K}(\gamma_0)$ only,
i.e., the $K$ largest estimated SNRs uniquely determines the BLER. 
Therefore, if the code rate $R$ is low (i.e., the number of $K$ is small)
and estimation accuracy is good for low SNR region, the estimated BLER 
of \eqref{eq:BLER} should become tight compared to the true BLER,
which may be estimated by time-consuming simulations with a sufficiently large number of trials.

\subsection{Comparison for Short Polar Codes}

We first compare the minimum estimated BLER based on~\eqref{eq:BLER}
calculated for the conventional GA, IGA of~\cite{ochiai2021}, and the proposed RCA 
as well as the corresponding simulation results in the cases of $N=64$ and $1024$ with 
three different code rates ($R=0.25, 0.5, 0.75$).
The results are listed in Table~\ref{table_designSNR}.
Note that the channel SNR (or design SNR) 
$\gamma_0$ is selected such that the resulting BLER reaches
around $10^{-2}$, and the same value of $\gamma_0$ is applied for all three cases.
Also, if the same polar code as IGA is constructed by 
any of the other schemes, this is indicated in the table. 
In such a case, the resulting code is the same and thus
the simulation result for the BLER should be 
identical to that reported in the IGA column.

Analogous to the observation in~\cite[Chap.~7.4.1]{chung_diss_2000}, 
the performances are similar for all the cases evaluated here.
Note that the polar codes constructed by the three different schemes
are identical in the case of $N=64$. With a length of $N=1024$
and for the cases of $R=0.25$ and $0.5$,
the codes constructed by RCA and IGA differ 
by only one bit position, 
and the corresponding simulation results indicate that the code designed by
RCA very slightly outperforms that of IGA, whereas all the codes
designed by the conventional GA are the same as those by IGA even in this case.

\begin{table*}[tbp]
 \begin{center}
  \caption{Comparison of estimated BLER and simulation.}
  \label{table_designSNR}
  \begin{tabular}{|c|c|c|c|c|c|c|c|c|} \hline
\multirow{2}{*}{$N$} & \multirow{2}{*}{$R$}  & \multirow{2}{*}{$\gamma_0$ [dB]}
 & \multicolumn{2}{c|}{BLER\,(RCA)} & \multicolumn{2}{c|}{BLER\,(IGA)} & 
\multicolumn{2}{c|}{BLER\,(GA)} \\ \cline{4-9}
& & & Estimate & Simulation
& Estimate & Simulation
& Estimate & Simulation \\ \hline
\multirow{3}{*}{$64$} & 
$0.25$ & 
$-2.53$ & 
$0.0100$ & = IGA
& $0.0100$ & $0.0095$ 
& $0.0101$ & = IGA  \\ \cline{2-9}
&
$0.5$ & 
$0.65$ & 
$0.0100$ & = IGA 
& $0.0100$ & $0.0096$ 
& $0.0102$ & = IGA \\   \cline{2-9}
&
$0.75$ & 
$3.26$ & 
$0.0101$ & = IGA 
& $0.0100$ & $0.0099$ 
& $0.0103$ & = IGA  \\ \hline
\multirow{3}{*}{$1024$} 
&
$0.25$ & 
$-3.97$ & 
$0.0103$ & $0.0100$ & $0.0111$ & $0.0102$ & $0.0113$ & = IGA  
\\ \cline{2-9}
&
$0.5$ & 
$-0.50$ & 
$0.0102$ & $0.0103$ & $0.0106$ & $0.0104$ & $0.0109$ & = IGA \\   \cline{2-9}
&
$0.75$ & $2.33$ & $0.0104$ & 
= IGA 
& $0.0102$ & $0.0111$ 
& $0.0109$ & = IGA
\\ \hline
 \end{tabular}
 \end{center}
\end{table*}

\begin{figure} [tbp]
\begin{center}
\includegraphics[width=.7\textwidth,clip]{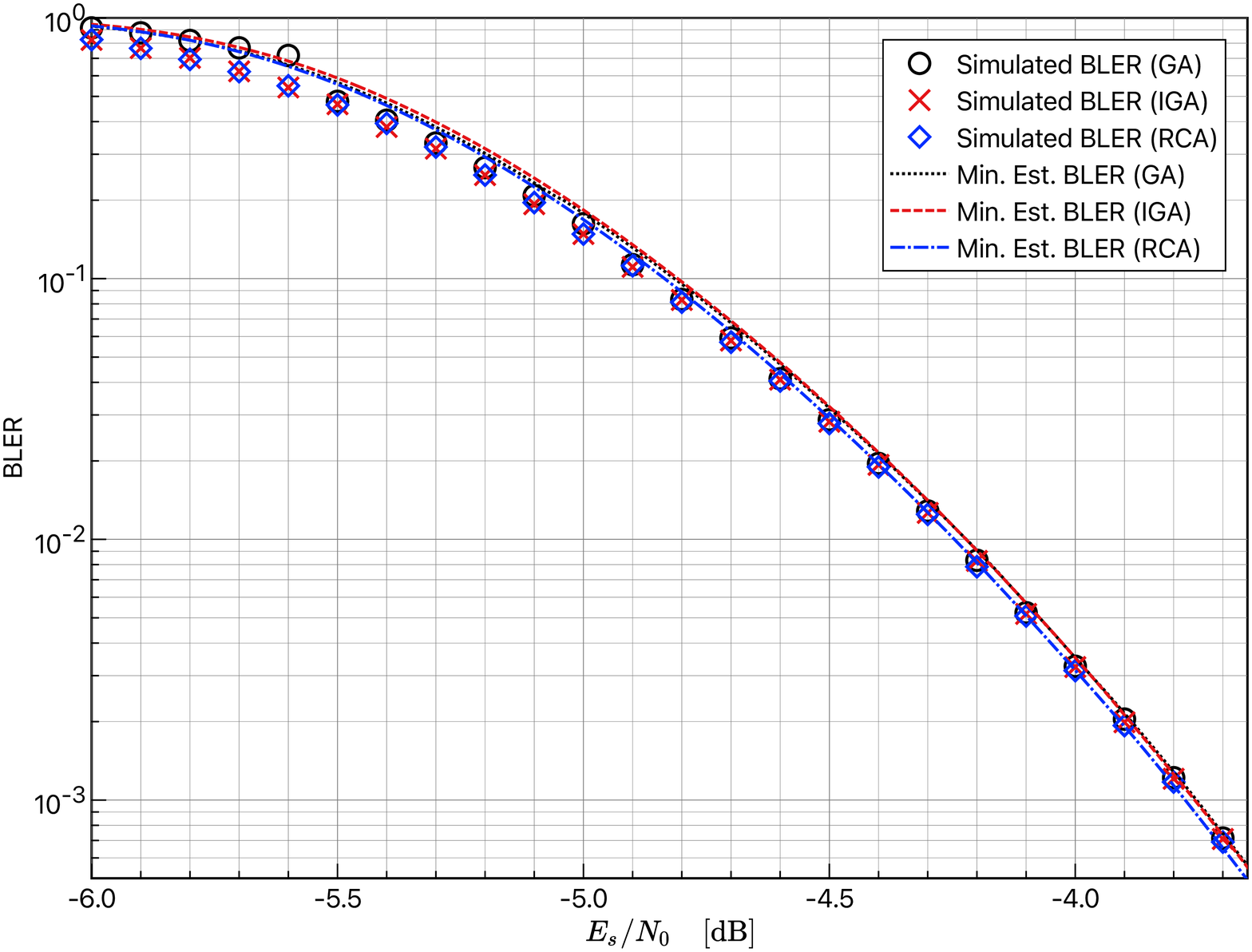}\\
\vspace{-.3cm}
(a)\\
\quad \vspace{-.3cm}\\
\includegraphics[width=.7\textwidth,clip]{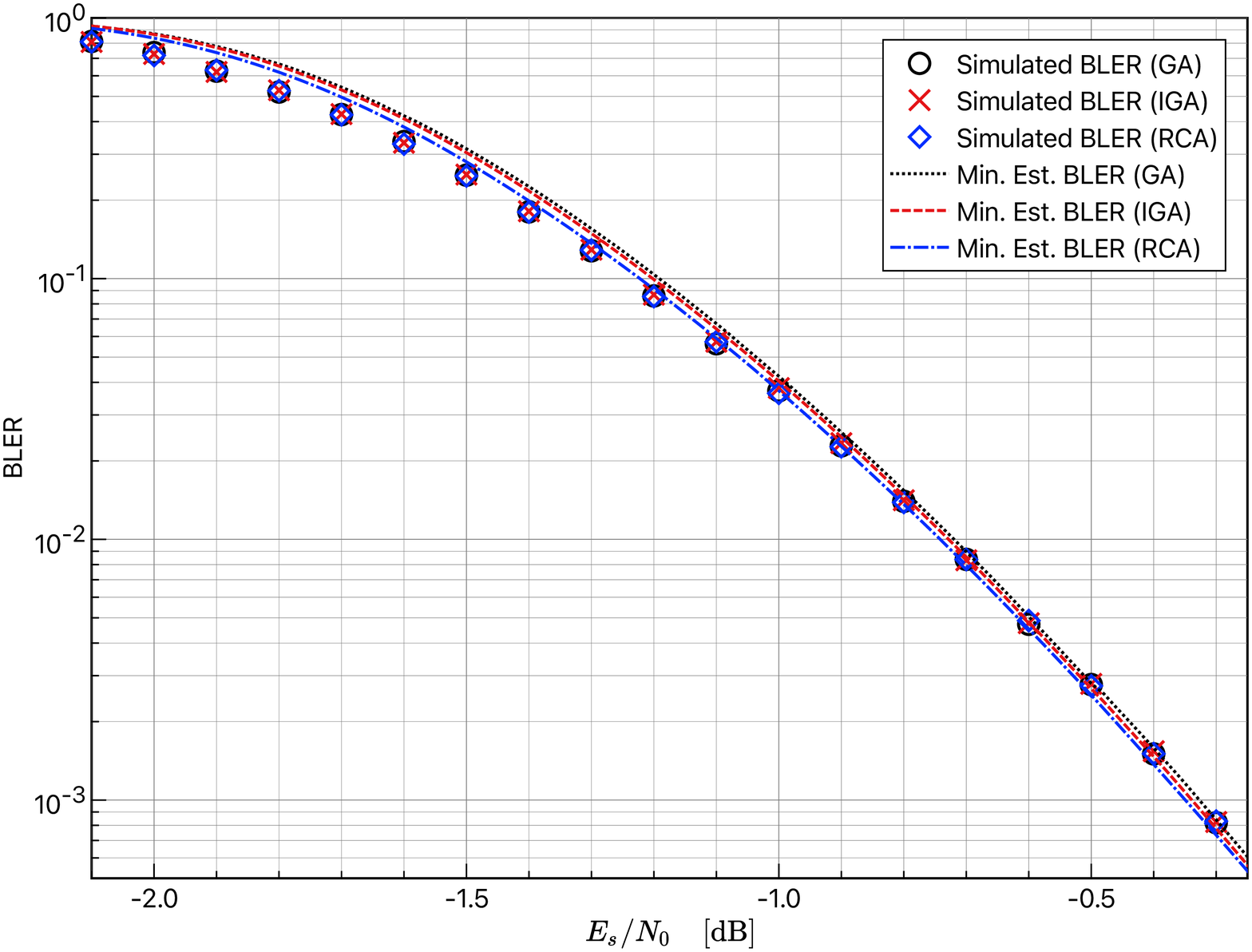}\\
\vspace{-.3cm}
(b)\\
 \caption{Simulated BLER of the polar codes with $N=2^{11}$ designed at each channel SNR
as well as the minimum estimated BLER: (a) $R=0.25$, (b) $R=0.5$.}
\label{fig:short_polar_codes}
\end{center}
\end{figure}

The BLER performances of these three approaches 
with $N=2048$ bits
are compared in Fig.~\ref{fig:short_polar_codes} 
for code rates of $R=0.25$ and $R=0.5$.
Note that for all the simulation results, 
the polar code is constructed {\em for each given channel SNR}.
We observe that in the case of low code 
rate ($R=0.25$) and low SNR region ($E_s/N_0 < -5.5\,{\rm dB}$), 
the performance degradation of GA compared to IGA and RCA is noticeable, 
as the approximation error of the conventional GA becomes significant 
with decreasing SNR as
discussed in~\cite{ochiai2021}. On the whole, however, 
the performance difference is almost negligible
similar to those reported in Table~\ref{table_designSNR},
justifying the effectiveness of the conventional GA as well as IGA
for the polar codes with modest codeword lengths.
Nevertheless, it is worth mentioning that the simulation results match
best with the minimum estimated BLER calculated by the proposed RCA algorithm.

\subsection{Comparison for Long Polar Codes}

\begin{figure} [tbp]
\begin{center}
\includegraphics[width=.7\textwidth,clip]{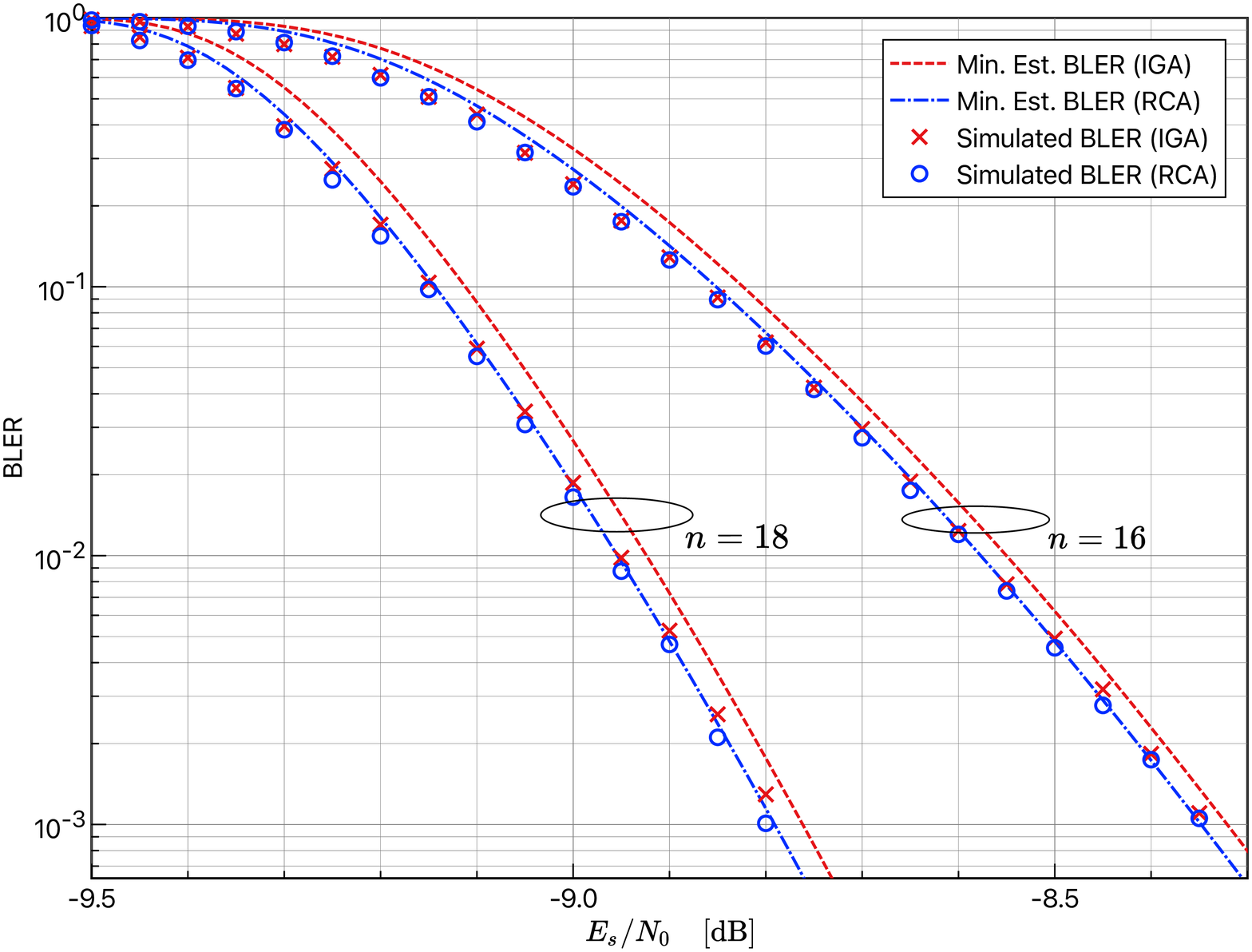}\\
\vspace{-.3cm}
(a)\\
\quad \vspace{-.3cm}\\
\includegraphics[width=.7\textwidth,clip]{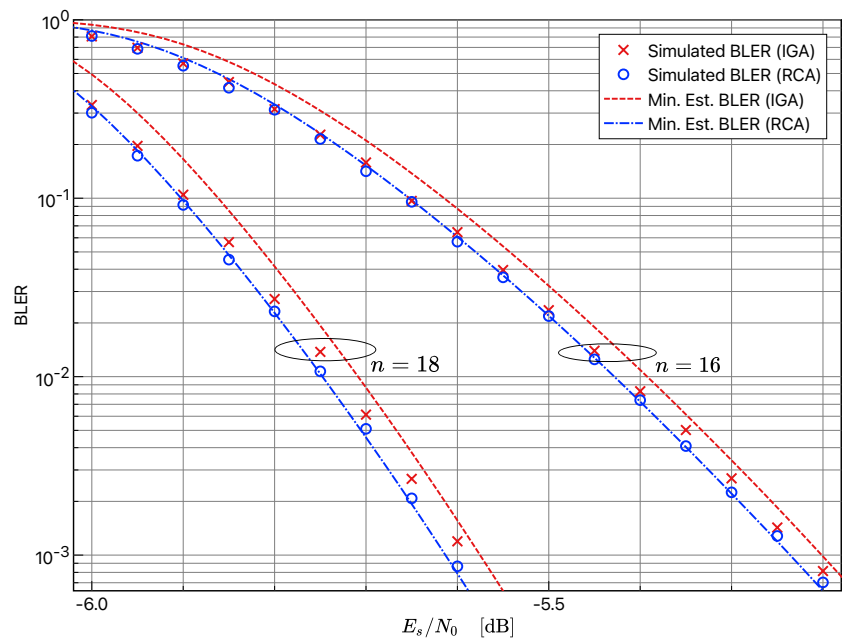}\\
\vspace{-.3cm}
(b)\\
 \caption{Simulated BLER of the long polar codes with $N=2^n$ 
($n=16$ and $18$) designed at each channel SNR
as well as the minimum estimated BLER: (a) $R=0.125$, (b) $R=0.25$.}
\label{fig:long_polar_codes_low}
\end{center}
\end{figure}

\begin{figure} [tbp]
\begin{center}
\includegraphics[width=.7\textwidth,clip]{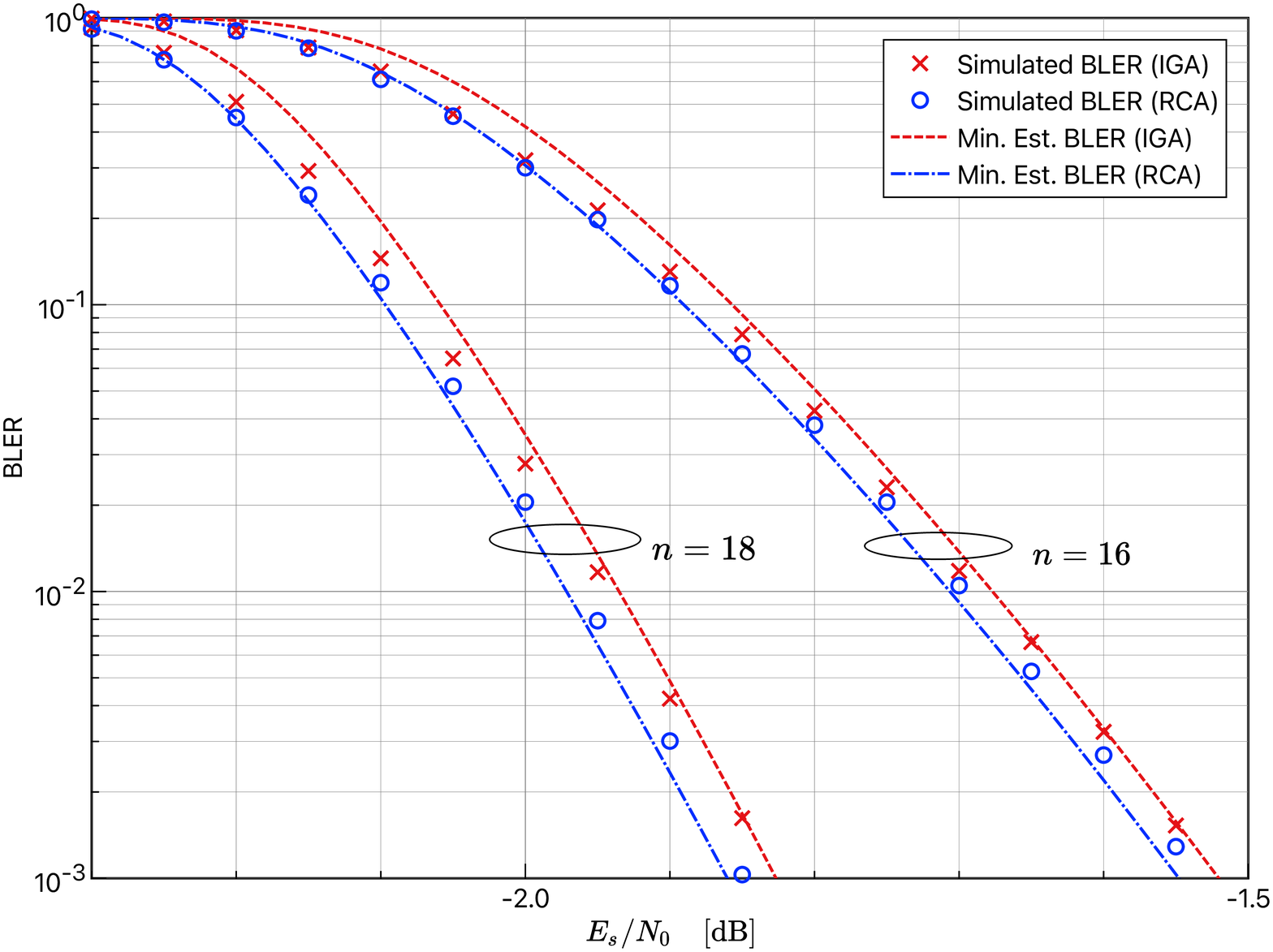}\\
\vspace{-.3cm}
(a)\\
\quad \vspace{-.3cm}\\
\includegraphics[width=.7\textwidth,clip]{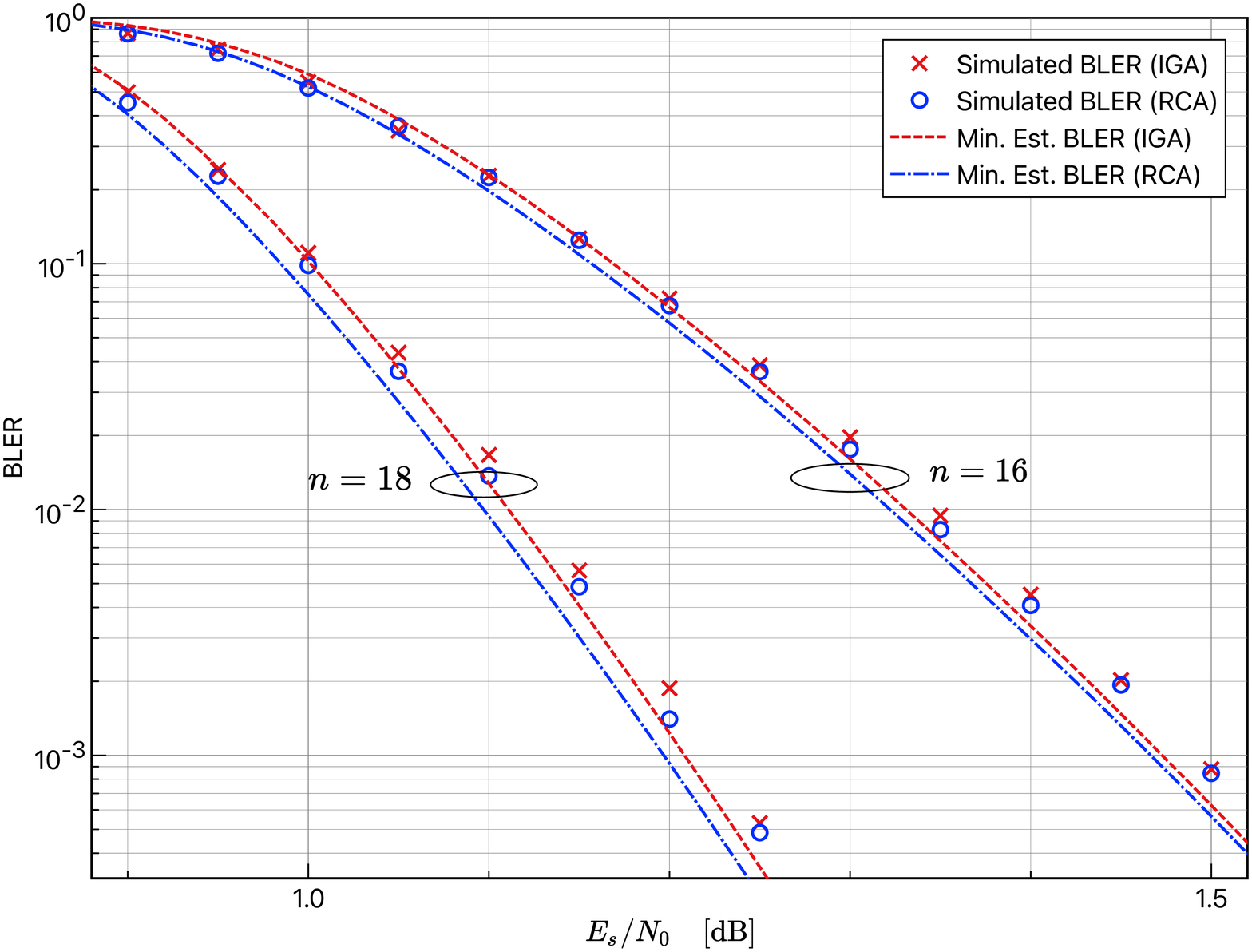}\\
\vspace{-.3cm}
(b)\\
 \caption{Simulated BLER of the long polar codes with $N=2^n$ ($n=16$ and $18$) designed at each channel SNR
as well as the minimum estimated BLER: (a) $R=0.5$, (b) $R=0.75$.}
\label{fig:long_polar_codes_high}
\end{center}
\end{figure}

We now focus on the polar code performance with long block lengths ($N=2^{16}$
and $2^{18}$).
Note that the performance of the conventional GA is not shown in the remaining figures
as it becomes significantly worse than that of IGA as demonstrated in \cite{ochiai2021}.
The cases with lower code rates ($R=1/8$ and $R=1/4$)
are compared in Fig.~\ref{fig:long_polar_codes_low},
whereas those with higher code rates ($R=1/2$ and $R=3/4$)
are shown in Fig.~\ref{fig:long_polar_codes_high}.
We observe that for all the cases compared, the polar codes
designed based on the proposed RCA algorithms outperform those based on IGA,
and the gap is especially noticeable when the code rate is low.
We also notice that, similar to the results in Fig.~\ref{fig:short_polar_codes}, 
the minimum estimated BLER values computed from~\eqref{eq:BLER}
and those obtained by simulation
are in general closer when RCA is employed, indicating that the estimation accuracy 
of the equivalent SNR obtained by RCA is higher than that of IGA.
Therefore, considering the fact that the computational complexity required for performing 
GA, IGA, and RCA is almost the same, the proposed RCA may be preferable to GA-based approaches, especially when polar codes with long block lengths and 
moderate code rates should be designed under severe limitation of computational resources.

\section{Conclusions}
\label{sec:conc}

In this work, we have proposed an explicit closed-form algorithm for polar code design based
on RCA. Simulation results have shown that polar code designed based on the proposed 
RCA can outperform popular Gaussian approximation~(GA)-based approaches with no increase of computational complexity. Moreover, the estimated BLER performances 
from RCA are closer
to those obtained by simulation compared to those based on GA, indicating the accuracy
of RCA compared to GA. 
The benefit of the proposed approach becomes 
significant for polar codes with large block lengths and 
moderate code rates.

\end{document}